\documentclass[twocolumn,amsmath,prb]{revtex4}

\usepackage[dvips]{color}
\usepackage{graphicx}% Include figure files
\usepackage{dcolumn}% Align table columns on decimal point
\usepackage{bm}% bold math
\def\be{\begin{equation}}
\def\ee{\end{equation}}
\def\x{\hat{\sigma_{x}}}
\def\z{\hat{\sigma_{z}}}
\def\adag{\hat{{a_{l}}}^{\dagger}}
\def\a{\hat{a_{l}}}
\def\freq{\omega_{l}}
\def\g{g_{l}}
\def\s{\sum_{{l}}}
\def\f{f_{l}}

\def\k{\tilde K}
\def\bra#1{\mathinner{\langle{#1}|}}
\def\ket#1{\mathinner{|{#1}\rangle}}

\def\p#1{#1^{'}}

\begin{document}

%\preprint{}

\title{Coherent-incoherent transition  in the sub-Ohmic spin-boson model}

\author{Alex Chin}
 \affiliation{Theory of Condensed Matter group, Cavendish Laboratory, University of Cambridge, Cambridge,
CB3 0HE, UK}
%Lines break automatically or can be forced with %\\
\author{Misha Turlakov}
\affiliation{Peierls Centre for Theoretical Physics, University of Oxford,
Oxford, OX1 3NP, UK}%

\date{\today}% It is always \today, today,

\begin{abstract}
We study the spin-boson model with a sub-Ohmic bath using a variational method.
The transition
from coherent dynamics to incoherent tunneling is found to be abrupt as a function of
the coupling strength
$\alpha$ and to exist for any power $0 < s< 1$, where the bath coupling is
described by
$J(\omega) \sim \alpha \omega^{s}$. We find  non-monotonic temperature
dependence
of the two-level gap $\tilde{K}$ and a re-entrance regime close to the
transition
due to non-adiabatic low-frequency bath modes. Differences between
thermodynamic and dynamic
conditions for the transition as well as the limitations of the simplified bath
description are discussed.
\end{abstract}

\maketitle

% ----------------------------------------------------------------
\section{Introduction}
The spin-boson model\cite{leggett1} is a paradigm model for the study of dissipation and decoherence in quantum
mechanics, and as such it is has been applied in
a wide range of systems. Such applications include the search for macroscopic
quantum coherence\cite{tagaki}, electron transfer in chemical\cite{egger} and
biological\cite{parson} physics, and most recently, the problem of dephasing
and relaxation in solid state qubits\cite{shnirmannoise}.

The particular case of the sub-Ohmic spin-boson model has had an interesting
development in the last few years. Compared
to the Ohmic bath, the sub-Ohmic bath is characterised by an increased density
of states for the low frequency bath modes.
This makes analysis of the dynamics difficult
as the low frequency modes generally lead to non-Markovian dynamics and strong
memory effects, even for relatively weak coupling.
One of the physical situations corresponding to the sub-Ohmic bath is the $1/f$
noise
in Josephson qubits\cite{shnirmannoise,paladino}. Although there are certain
limitations and assumptions in describing
$1/f$ noise by an equilibrium sub-Ohmic bath\footnote{These limitations have to do
with the weakness of coupling to each individual bath mode as well as
relaxation times for bath modes to thermal equilibrium. See also Y.M. Galperin,
B.L.Altshuler, D.V. Shantsev, in "Fundamental Problems of Mesoscopic Physics",
Eds. I. V. Lerner et al. (Kluwer Academic Publishers, The Netherlands, 2004),
pp.141-165.}, the study of the sub-Ohmic spin-boson model
may be useful in this and other contexts.

In the earliest treatments of the sub-Ohmic model\cite{leggett1}, it was argued
that the sub-Ohmic bath
always destroys the coherence of superposition states and localises the system
in one state for any non-zero coupling.
This conclusion was based on the non-interacting-blip
approximation\cite{leggett1,weiss} which fails
in the weak-coupling limit to the bath.
More recently, several works\cite{kehrein,shnirmannoise,bulla} addressed the
problem of the sub-Ohmic spin-boson model and found that coherent phases can exist for sufficiently weak coupling.

In the light of these developments, we contribute to this discussion of the coherent dynamics of the
sub-ohmic model
by demonstrating the existence of the coherent regime for arbitrary $s<1$
using a simple and intuitive variational method. The variational method we use
was originally developed by Silbey and Harris \cite{silbey} for the problem of
Ohmic damping. We will show
that such a treatment allows us to define precise criteria for the thermodynamic existence of a
coherent phase, and also provides a means to quantitatively map out the
parameter space of the sub-Ohmic coherent regime. Within the coherent regime,
we also give new results for the renormalisation of the parameters that
describe the coherent dynamics.
A re-entrance regime close to the coherent-incoherent transition is found. In
addition, strong coupling to non-adiabatic
modes is considered, showing the limitations of Silbey-Harris variational
ansatz.

In section \ref{sb} we briefly give an outline of the spin-boson model and in
section \ref{var} we describe the variational  method and explain the simple
physical picture behind the variational ansatz. Throughout this paper we are
primarily concerned with finding the conditions under which coherent
oscillations of the TLS are possible, and in sections \ref{o} and \ref{T} we
give some new quantitative results for the critical couplings and renormalised
parameters of the dissipative tunneling at zero and finite temperature. In
sections \ref{dynamics} and \ref{correction}, we highlight some of the limitations of this method and in
section \ref{disc} we discuss our results and compare the findings with the results obtained by other authors. We end with a brief
conclusion and summary.

\subsection{The spin-boson model}
\label{sb}
The spin-boson model consists of a single two-level system (TLS) coupled
linearly to an infinite bath of harmonic oscillators.
The TLS can be thought of as spanning the two lowest levels of a double well
potential, or
in other contexts, the TLS may appear in the situation where the transition matrix
element between two given energy levels
is much larger than the transition matrix elements to all other energy levels of
the system.
The two levels are coupled by a tunneling matrix element $K$, and taking these
levels to be eigenstates of $\z$,
the spin-boson Hamiltonian is given by,

\begin{equation}
\hat{H}= K\x + \epsilon\z + \s\freq(\adag\a +1/2) +\z\s\g(\adag+\a).
\label{sbh}
\end{equation}
$\epsilon$ is the bias energy between the minima of the wells and for the rest
of this paper, we set $\epsilon=0$.
$\a,\adag$ are the bosonic annihilation/creation operations for the bath modes.
The last term in Eq(\ref{sbh}), the linear coupling of $\z$
to the coordinate displacement of the bath oscillators, is assumed.

In the absence of coupling, the eigenstates are coherent superpositions of the
left and right states, and the particle can oscillate in time between the wells
at a frequency $2K$. The couplings $\{\g\}$ between the spin (or TLS)
 and the oscillators generally lead to damping of this motion, or may even
suppress tunneling entirely. For an in-depth discussion of the rich dynamics
of spin-boson models, the reader is referred to the original review of Leggett
et al\cite{leggett1} or the more recent collection of papers on this
subject\cite{grabert}.
The observed dynamical behaviour is determined by the spectral function of the
bath, $J(\omega)$,
\begin{equation}
J(\omega)=\pi\s\g^{2}\delta(\omega-\freq).
\end{equation}
For frequencies below a high energy cut-off, $\omega_{c}$, the spectral
function can be modeled by the power law form,
\begin{equation}
J(\omega)=\frac{1}{2}\pi\alpha\omega_{s}^{1-s}\omega^{s}
\label{spec}
\end{equation}
where $\alpha$ is a dimensionless parameter that measures the effective
strength of the system-bath coupling and $\omega_{s}$ is an energy scaled
included to keep $\alpha$ dimensionless. In this paper we assume that the cut-off frequency $\omega_{c}$ is much greater than all other scales in the problem, but the scale $\omega_{s}$ can be large or small compared to other scales e.g. $K,\,K_{B}T$.

The case of a bath with $s=1$ is known as the Ohmic bath, and the dynamics and
thermodynamics of this model have been studied extensively in the
literatue\cite{leggett1,weiss}. Baths described by $s>1$ are termed super-Ohmic
and we will not discuss them further. Here we shall focus on the sub-Ohmic spin-boson model where the bath is characterised by $0<s<1$.

\section{The Variational Method}
\label{var}
In this section, we motivate the variational approach to this problem
\cite{silbey} and outline the method.
In the absence of the tunneling term, the spin-boson Hamiltonian Eq.(\ref{sbh})
reduces to the well-known independent boson model\cite{mahan}, and
the solutions of Eq.(\ref{sbh}) correspond to the particle localised in one of the
wells. The oscillator part of the Hamiltonian is then just a collection of
displaced oscillators and can be diagonalised by a simple translation of the
oscillators $\propto\g \langle\z \rangle$.

If the tunneling is switched back on we propose that the approximate
eigenstates of the system correspond to a dressed particle tunneling between
the two levels carrying a dynamical cloud of such oscillator displacements. Our
variational ansatz is
therefore in the spirit of an adiabatic approximation\cite{leggett1} but with a special
treatment of  the low frequency,
non adiabatic modes as we discus further in section \ref{disc}.

To describe this physical picture, we reproduce below the technique
originally used by Silbey and Harris\cite{silbey} for the ohmic bath. We begin by
performing a unitary transformation of the spin-boson Hamiltonian Eq.(\ref{sbh})
\begin{equation}
\tilde{\hat{H}}=\hat{U}\hat{H}\hat{U}^{-1}
\label{H}
\end{equation}
The unitary operator $U$ is given by,
\begin{equation}
\hat{U}=\exp\Bigl[\,-\z\,\s\,\freq^{-1}\,\f\,(\a -\adag)\Bigr].
\label{U}
\end{equation}
The arbitrary coupling parameters ${\f}$ introduced in Eq.(\ref{U}) are
proportional to the effective displacement or dressing of each bath mode due to
the coupling to the TLS. If $\f=\g$ then the transformation diagonalises the
last three terms of Eq.(\ref{sbh}), but as we demonstrate in section (\ref{disc}),
setting $\f=\g$ is often a sub-optimal choice for $\{\f\}$.

As we mentioned in the introduction, we are primarily interested in
establishing whether or not coherent oscillations can exist in the sub-Ohmic
model and to answer this question, we introduce the quantity $\k$ which is
given by,
\begin{equation}
\tilde K = K\,\Big\langle \exp\{ -2\s\,\f\,\freq^{-1}(\a -\adag)\}
\Big\rangle_{bath},
\label{k}
\end{equation}

where the brackets denote the thermal expectation value taken over the bath modes.

We interpret $\k$ as the effective coherent tunneling matrix element of the
dressed particle. The exponential factor in Eq.(\ref{k}) suppresses the bare tunneling, and this factor arises due to the partial overlap of the oscillators that dress the TLS as it tunnels between the wells. Adding
and subtracting $\k$ to Eq.(\ref{H}), we re-write the Hamiltonian as

\begin{equation}
\tilde{\hat{H}}=\hat{H_{0}}+\hat{V}
\label{ham}
\end{equation}
Where the separation is into a main coherent part $H_{0}$,
\begin{eqnarray}
\hat{H_{0}}&=&\tilde K\x+\s\freq(\adag\,\a +1/2)\label{ho}\\
&+&\s(\f^{2} -2\,\f\,\g)\nonumber,
\end{eqnarray}
and a series of perturbation terms, $\hat{V}$, which contain the remaining weak
coupling between the TLS and the bath,
\begin{eqnarray}
\hat{V}&=&\hat{V}_{+}\hat{\sigma}_{+} +\hat{V}_{-}\,\hat{\sigma}_{-} + \hat{V}_{0}\,\z,\\
\hat{V_{0}}&=&\s\,(\g-\f)\,(\a +\adag)\label{dep}\\
\hat{V_{+}}&=&\hat{V}_{-}^{*}=K\exp\{ -2\s\,\f\,\freq^{-1}(\a -\adag)\} -\k.
\label{V}
\end{eqnarray}

The introduction of $\k$ and the separation of the Hamiltonian into a coherent part and perturbations is reminiscent of a mean-field type theory, and our treatment is essentially in this spirit. Considering the main part of the Hamiltonian, Eq.(\ref{ho}), we see that if $\k$ is finite, the eigenstates of are coherent superpositions of the two levels and the TLS can undergo coherent oscillations between it's levels. If $\k$ vanishes, the degenerate levels become uncoupled and no coherent oscillations are possible. Therefore, at the mean field level, we can use the existence of a finite effective tunneling matrix element as the signature for the existence of a thermodynamic coherent phase in the TLS. The point where $\k$ vanishes as a function of the system parameters marks the transition to the incoherent phase.

However, this thermodynamic criterion for distinguishing between coherent and
incoherent dynamics is only approximate, as we have not yet included the dynamical effects
of the perturbation terms on the TLS dynamics. The effect of the perturbations
and the general limitations of using $\k$ as the criteria for the transition
from coherence to incoherence will be discussed in section (\ref{dynamics}).

To calculate $\k$, we must first determine the $\{\f\}$. Following Silbey and
Harris \cite{silbey}, we compute the Bogoliubov-Feynman\cite{feynmanstat} upper
bound on the free energy of the system, $A_{B}$\cite{feynmanstat}. Bogoliubov's
theorem\cite{mazo} states that the true free energy, $A$, of the Hamiltonian
(\ref{ham}) is related to $A_{B}$ by,
\begin{equation}
A \leq A_{B}\nonumber
\end{equation}
\begin{equation}
A_{B} = -\beta^{-1}\,\ln \textrm{Tr}\exp(-\beta\,\hat{H_{0}}) + \langle
\hat{V} \rangle _{H_{0}} + O(\langle\hat{V}^{2}\rangle_{H_{0}}).
\label{ab}
\end{equation}

Due to our choice of $H_{0}$, we have constructed the perturbations so
that $\langle V_{i}\rangle_{H_{0}}=0$. In section \ref{correction} we
explicitly calculate the second order terms and find that they give a small
contribution to $A_{B}$ if $\omega_c \rightarrow \infty$. Therefore, dropping all
higher-order terms, $A_{B}$ is given by,

\begin{equation}
A_{B}=-K_{B}T\ln[2\cosh(\k\beta)] + \sum_{l}\omega_{l}^{-1}(\f^{2}-2\f\g),
\label{a}
\end{equation}
where we left out a term due to the free bath ground state energy which does
not
depend on  $\{\f\}$. $A_{B}$ can then be minimised by varying the $\{\f\}$ to find,

\begin{equation}
\f=\g\,\left(1+2\,\tilde
K\,\freq^{-1}\,\coth(\freq\,\beta/2)\,\tanh\beta\,\tilde K\right)^{-1}.
\label{f}
\end{equation}

Notice already at this stage the limiting behaviour of coefficients $\{\f\}$:

\begin{equation}
\f \approx \left\{ \begin{array}{ll}
                       \g  & \mbox{if $\freq \beta \gg 1$ and $\tilde{K} \beta
\ll 1$}  \\
                       \g \frac{\freq}{2\tilde{K}}  & \mbox{if $\freq \beta \gg
1$ and $\freq \ll \tilde{K}$}
                    \end{array}
            \right.
\end{equation}
The coefficients of effective coupling  $\{\f\}$ vanish in the limit
$\omega_{l} \rightarrow 0$ and finite $\tilde{K}$.

We now substitute this form for $\{\f\}$ back into Eq.(\ref{k}) and
use the spectral function, Eq.(\ref{spec}), to turn the sum over modes in
Eq.(\ref{k}) into an integral. We then obtain our key equation, the self-consistent
equation for $\tilde{K}$

\begin{equation}
\tilde K=K\exp(-2F[\k])
\label{Kspec}
\end{equation}
\begin{equation}
F[\k]=\frac{1}{\pi}\int _{0}^{\omega_{c}}\frac {J(\omega)\, \coth
(\frac{\omega\,\beta}{2})\,d\omega}{ \big(\omega+2\,\k\tanh (\beta\,\k)
\coth (\frac{\omega\,\beta}{2}) \big) ^{2}}.
\label{F}
\end{equation}

\section{results at $T=0$}
\label{o}

The appearance of $\k$ on both sides of Eq.(\ref{Kspec}) means that
we must solve self-consistently for $\k$. At $T=0$, we can
perform the integral in Eq.(\ref{Kspec}) exactly. The results below were
calculated by extending the upper limit in Eq.(\ref{F}) to infinity, an
approximation that be easily dropped but which is a possible approximation for
$s<1$. Calculating the integral and substituting into Eq.(\ref{Kspec}), we obtain,

\begin{equation}
\k\exp\left(\frac{\alpha\omega_{s}^{1-s}\pi\,s}{(2\k)^{1-s} \sin(\pi s)}\right)=K.
\label{graph}
\end{equation}

The self-consistent values of $\k$ can then be obtained by numerically solving
Eq.(\ref{graph}) for general values of $\alpha$. Note that $\k=0$ is always a solution of Eq.(\ref{graph}).

Using Eq.(\ref{graph}), it is possible to determine the critical coupling strength
$\alpha_{c}$ for fixed $K$. $\alpha_{c}$ is the coupling strength above which
the only possible solution of Eq.(\ref{graph}) is $\k=0$.

\begin{figure}
\includegraphics[width=8.6cm]{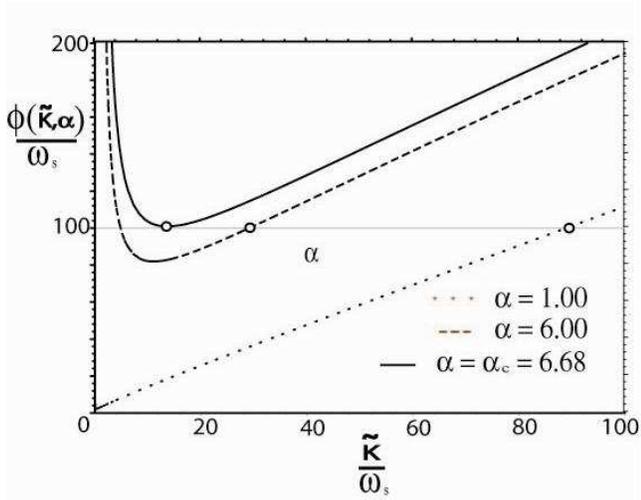}
\caption{Solutions of the self-consistent equation for $s=1/2$ and $K/\omega_{s}=100$. Solutions exist where $\phi(\k)$ intersects the line $\k=K$. As the coupling $\alpha$ is increased, the curve shifts until there is no intersection. The coupling where this occurs defines the critical coupling $\alpha_{c}$.}
\label{graphical}
\end{figure}

To see the existence of this critical coupling, we define the LHS of Eq.(\ref{graph}) as $\phi(\alpha,\k)$. This function has the typical
form shown in Fig.\ref{graphical}, and crucially, has only one minimum for any sub-Ohmic bath. Finite solutions for $\k$ exist when the
$\phi(\alpha,\k)$ intersects with the line $\k=K$, and the point of intersection is controlled by the coupling strength $\alpha$ as shown in Fig.(\ref{graphical}). The critical coupling
strength can then be clearly identified as the coupling strength where the minimum of $\phi(\alpha,\k)$ just touches the line
$\k=K$ as shown in Fig.(\ref{graphical}). When $\alpha>\alpha_{c}$ there is
no longer any intersection with the line $\k=K$ and the only self-consistent
value of the renormalised tunneling matrix element is $\k=0$.

The position and value of the minimum in $\phi(\k)$ can be determined by
elementary calculus, and this gives the  results,

\begin{eqnarray}
2\k_{min}&=&\omega_{s}\left(\frac{\alpha_{c}\Delta(s)}{e}\right)^{1/(1-s)}\\\\\nonumber
\phi(\k_{min})&=&\k_{min}e^{(\frac{1}{1-s})}=K,
\end{eqnarray}
where $\Delta(s)=e\pi\,s(1-s)/\sin(\pi s)$. From these equations we then find
that the critical coupling $\alpha_{c}$ is given by,
\begin{equation}
\alpha_{c}=\frac{1}{\Delta(s)}\left(\frac{2K}{\omega_{s}}\right)^{1-s}.
\label{zero}
\end{equation}
Note that unlike the case of ohmic damping, the sub-Ohmic critical coupling depends on
the ratio of $2K/\omega_{s}$, and that the Silbey-Harris approach predicts a finite $\alpha_{c}$ as $s \rightarrow 0$.
Note also that the above condition (Eq.\ref{zero}) can be rewritten so it is a condition
on the coefficient $\alpha \omega_s^{1-s}$ appearing in the spectral function (Eq.\ref{spec}).

We also find that when $\alpha<\alpha_{c}$, $\k$ satisfies,
\begin{equation}
K\exp\{\frac{1}{s-1}\}\leq \k\leq K \hspace{1cm}\mathrm{if}\hspace{0.5cm}
\alpha<\alpha_{c}.
\label{ineq}
\end{equation}
This inequality shows that at $T=0$, $\k$ undergoes a discontinuous
jump from a finite $\k$ to $\k=0$ as $\alpha\rightarrow\alpha_{c}$ Only at
$s=1$ does this method predict that $\k\rightarrow 0$ continuously.

In some other treatments of this problem\cite{bulla}, the energy scale
$\omega_{s}$ is set equal to the high frequency cut-off $\omega_{c}$. If this
is done in this method, we find that we cannot send
$\omega_{c}\rightarrow\infty$ in Eq.(\ref{F}) as this leads to $\k=0$
for all $s$ and $\alpha$. Keeping $\omega_{c}=\omega_{s}$ finite, we get the
result,
\begin{equation}
\alpha_{c}e^{-s\alpha_{c}}=\frac{1}{\Delta(s)}\left(\frac{2K}{\omega_{c}}\right)^{1-s}.
\label{finite}
\end{equation}

The only modification to the previous result, Eq.(\ref{zero}), is the exponential
factor and the replacement $\omega_{s}\rightarrow\omega_{c}$. The exponential
factor is the correction for a finite cut-off and is important only as
$s\rightarrow 1$. Note that as $s\rightarrow 1$, Eq.(\ref{finite}) correctly
predicts the critical coupling for the ohmic case,
$\alpha_{c}=1$.\cite{leggett1}. We will not be too interested in the ohmic case
as this has already been thoroughly dealt with in the literature. Therefore for
the rest of the this paper we work with $s<1$, $\omega_{s}\neq\omega_{c}$ and
$\omega_{c}\rightarrow\infty$ in the integral Eq.(\ref{F}).

\begin{figure}
\includegraphics[width=8.6cm]{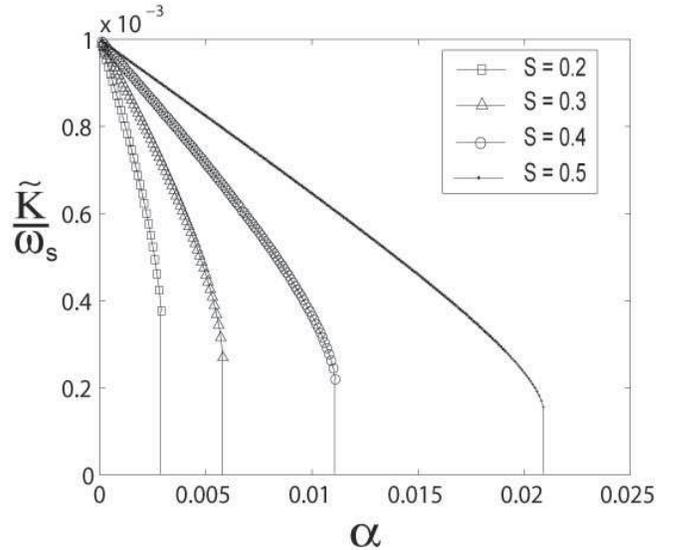}
\caption{Behaviour of $\k(\alpha)$ as a function of $\alpha$ at $T=0$. For this computation $K=10^{-3}\omega_{s}$ and $\omega_{c}\rightarrow \infty$.}
\label{zerograph}
\end{figure}

For $\alpha\ll 1$ we can make a perturbative expansion in $\alpha$ and
determine $\k(\alpha)$ to first order in $\alpha$,
\begin{equation}
\k(\alpha)=K\left[1-\frac{\alpha\,s\pi}{\sin(\pi s)}
\left(\frac{\omega_{s}}{2K}\right)^{1-s}\right].
\label{ko}
\end{equation}

For general couplings to the bath, equation (\ref{Kspec}) needs to be solved
numerically for $\k$, and the typical behaviour of $\k(\alpha)$ across the whole range of coupling strengths is shown in Fig.(\ref{zerograph}).

\section{Finite temperatures}
\label{T}
\subsection{High temperatures}
Calculating the integral in Eq.(\ref{F}) and solving Eq.(\ref{Kspec}) for the general
case of finite temperatures can only be done numerically. However some
analytical results can be extracted in certain limits. For the case of high
temperatures and very weak coupling where $\k\,\beta\ll\,1$, we again find a
coherent regime which crosses over to incoherent relaxation at a critical
temperature $T^{*}$,
\begin{equation}
T^{*}=\frac{K}{\alpha\,f(s)}\left(\frac{2K}{\omega_{s}}\right)^{1-s}
\hspace{1cm} \alpha < \alpha_{c},
\label{tstar}
\end{equation}
where $f(s)$ is a slowly varying function of $s$ which is always $\approx
O(1)$. In this regime, we find that the transition from finite $\k$ to $\k=0$
occurs discontinuously at $T^{*}$.

For stronger coupling the relation given by Eq.(\ref{tstar}) is violated,
and a numerical study we have performed shows that $\k$ vanishes at a
significantly lower temperature than $T^{*}$ as $\alpha\rightarrow\alpha_{c}$.
For weak coupling, the numerical calculations of $\k(T)$ give values of $T^{*}$
in good agreement with Eq.(\ref{tstar}).

\subsection{Low temperatures}
For temperatures close to zero where $\k\beta\gg1$, we can solve the
self-consistent Eq.(\ref{Kspec}) for weak coupling by making a
perturbation expansion in powers of $\alpha$. The result to first order is,
\begin{equation}
\k(T,\alpha)=\k(0,\alpha)+2\alpha\,g(s)\frac{(K_{B}T)^{2}}{K}\left(\frac{\omega_{s}}{2K_{B}T}\right)^{1-s},
\label{low}
\end{equation}
where $\k(0)$ is given by Eq.(\ref{ko}) and $g(s)$ is another function
of $s$ which is of order unity.
Eq.(\ref{low}) shows the surprising result that $\k$ becomes larger as
the temperature is increased from zero. This result was also derived by
Weiss\cite{weiss} for the ohmic bath and was qualitatively described by Kehrein
and Mielke\cite{kehrein} for the sub-ohmic bath. However we believe the
quantitative result given in Eq.(\ref{low}) have not been explicitly
presented before for the sub-ohmic bath. We shall discuss this effect in more
detail in section \ref{disc}.
\subsection{Intermediate temperatures}
\begin{figure}
\includegraphics[width=8.6cm]{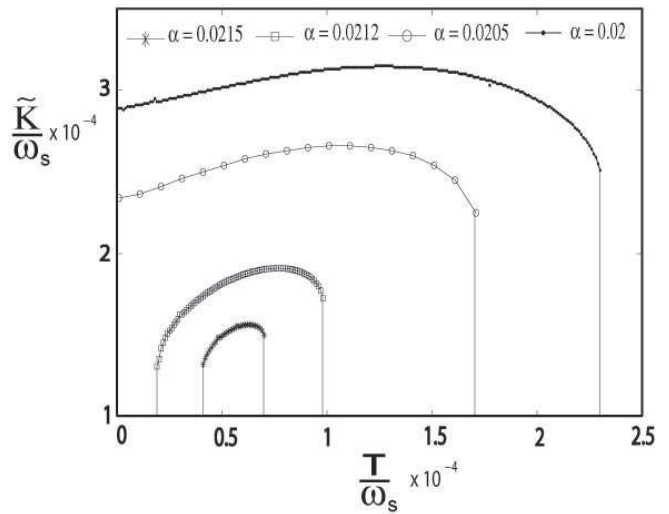}
\caption{$\k$ as a function of temperature for a sub-ohmic bath with $s=1/2$ and a range of different couplings. For this numerical computation  $K=10^{-3}\omega_{s}$ and $\omega_{c}\rightarrow \infty$ Note that the two lowest curves have finite values of $\k$ only between $T^{*}$ and a lower re-entrance temperature. For these curves, $\alpha>\alpha_{c}\approx0.021$. }
\label{temp}
\end{figure}

For intermediate values of $\alpha$ and $T$, we can determine $\k$ numerically
and the typical behaviour is shown in Fig.(\ref{temp}). In all cases we find that
$\k$ increases to a maximum and then drops discontinuously to $\k=0$ at
$T^{*}$.  In section \ref{disc} we estimate that the peak in $\k(T)$ should
occur approximately  at a temperature $K_{B}T_{max}\sim\k(T_{max})$, which for
sufficiently weak coupling can be approximated as $K_{B}T_{max}\sim K$.
Comparing to the numerical results we find that this is a good order of
magnitude estimate, but the peak typically occurs at a lower temperature $\sim
T_{max}/2$ as illustrated in Fig.(\ref{temp}).

 The numerical results also reveal an interesting feature if we look at the temperature dependence of $\k$ for systems with $\alpha>\alpha_{c}$. We find that for couplings slightly above $\alpha_{c}$, the TLS is incoherent at $T=0$, but then develops a finite $\k$ between some re-entrance temperature and $T^{*}$. In this re-entrance regime, $\k$ shows the same non-monotonic temperature dependence described above, and some examples of the behaviour of $\k$ in this ``super-critical regime" are shown in Fig.(\ref{temp}).

\begin{figure}
\includegraphics[width=8.6cm]{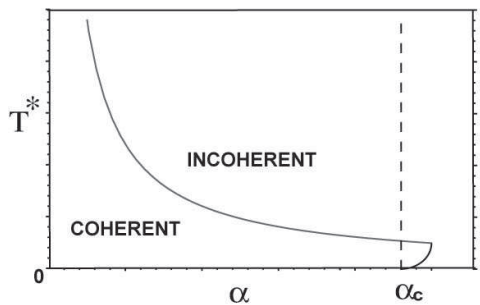}
\caption{Schematic plot of the boundary between coherent and incoherent regimes as found by the variational method. The curve is proportional to $1/\alpha$ for small $\alpha$. The plot also shows the small re-entrant region for coupling strengths greater than $\alpha_{c}$.}
\label{tstaralph}
\end{figure}

 As the coupling between the TLS and bath is increased, the re-entrance temperatures and $T^{*}$ merge to one finite temperature and beyond this coupling,
$\k=0$ for all temperatures. This region is generally very small and together
with the results for $\alpha<\alpha_{c}$, we obtain the schematic
coherent and incoherent regions of the sub-ohmic model as shown in Fig.(\ref{tstaralph}). This re-entrance phenomena is a consequence of the same mechanism that causes the enhancement of $\k$ at low temperatures, and we discuss this effect in section \ref{disc}.


\section{TLS dynamics and limitations of the variational method}
\label{dynamics}
In section \ref{var} we defined the criteria for coherent dynamics as the
existence of a finite renormalised tunneling matrix element $\k$. However this criterion
does not take into account the effect of the perturbation terms given in
equation (\ref{V}). These perturbation terms introduce dissipative dynamical effects which
can alter the oscillatory behaviour of the TLS in the coherent tunneling
state. These effects can be calculated by a variety of
methods\cite{leggett1,shnirmannoise,weiss,silbey} and for the ohmic case are well
understood.

For the sub-ohmic problem, we are interested in the weak-coupling
behaviour where approximations like the non-interacting blip
model\cite{leggett1,shnirmannoise} are no longer valid. The weak-coupling should
however permit us to analyse the effects of perturbations using the perturbative
reduced density matrix method.\cite{feynmanstat,blum}
To second order in the
perturbations, the reduced density matrix of the TLS, $\rho_{s}(t)$,is given by,

\begin{equation}
\dot{\rho}_{s}(t)=-\int^{t}_{0}d\p{t}\mathrm{Tr}_{b}\left[\hat{V}(t),\left[\hat{V}(\p{t}),\rho_{s}(\p{t})\rho_{b}(0)\right]\right],
\label{dmatrix}
\end{equation}
where the operators are written in the interaction representation
$\hat{V}(t)=\exp(i\hat{H_{0}}t)\hat{V}\exp(-i\hat{H_{0}}t)$,
and $\hat{V}$ and $\hat{H_{0}}$ are defined in Eq.(\ref{V}). $\rho_{b}(0)$ is the thermal density matrix for the unperturbed bath modes. Once $\rho_{s}(t)$ is known, all the observables of the TLS can be found using $\langle\hat{o}\rangle=\textrm{Tr}\,[\rho_{s}(t)\hat{O}]$, where the trace is only over the states of the TLS.

As can be seen in equation (\ref{dmatrix}), the time development of the reduced
density matrix depends on the whole history of it's motion, and such memory effects can lead to strong modification of the tunneling dynamics\cite{grabert}. As a simple example of the dynamical effects that perturbation can cause, we consider very weak coupling and ignore the memory structure of the bath. This simplification is known as the Born-Markov approximtion\cite{blum}, and applying it to the spin-boson model, we find that
$\z$ obeys the simple equation of motion,\cite{leggett1,silbey}
\begin{equation}
\frac{d^{\,2}\langle\z (t)\rangle}{dt^{2}}+2\Gamma\,\frac{d\langle\z (t)\rangle}{dt}+4K^{2}\langle\z(t)\rangle=0.
\label{eqm}
\end{equation}
Therefore, in the Born-Markov approximation, the coherent oscillations of the TLS are exponentially damped with a decay rate given by,
\begin{equation}
\Gamma_{osc}=J(2\k)\coth(\k\beta)
\label{gam}
\end{equation}
so that as $t\rightarrow\infty$ the TLS settles into a decoherent mixture of localised states. The coherence of the initial state is gradually destroyed by interactions with the environment on a time scale $1/\Gamma$. This is a generic phenomena for open quantum systems,\cite{blum} and for long enough times, the initial coherence of the superposition state is destroyed. Therefore when we talk about the coherent phase in the Silbey -Harris variational method, we mean that the initial ground state is coherent; the subsequent tunneling is then subject to decoherent and dissipative processes which eventually destroy the coherence.

The purpose of these remarks on dynamics is to point out that in the thermodynamic coherent phase, these decoherent and dissipative processes can potentially drive the coherent tunneling of the TLS to become incoherent. Therefore it is possible that there is a transition to incoherent motion due to dynamical effects that may occur before or after the thermodynamic transition we have found at $\alpha_{c}$ or $T^{*}$. For example, the Silbey-Harris variational method predicts $\alpha_{c}=1$ for the ohmic bath, whilst it is well known that for ohmic
baths at $T=0$,
there is localisation for $\alpha\geq 1$, incoherent tunneling for
$0.5<\alpha<1$, and coherent oscillations are only observed for $\alpha<0.5$.
\cite{leggett1}

In order to find such a dynamical cross-over in the variational method, we need to account for the perturbation terms and solve for the dynamics of the TLS. For Ohmic damping, Silbey and Harris calculated that the crossover to incoherent tunneling occurs when the equation of motion (\ref{eqm}) becomes overdamped, which occurs at a coupling strength $\alpha_{inc}=2/\pi$ at $T=0$. If we apply the same procedure to the sub-Ohmic bath then we find that,
\begin{equation}
1= \frac{\alpha_{inc} \pi}{2} \left( \frac{2K}{\omega_s} \right)^{s-1} coth(\beta K)
\end{equation}
after substituting Eq.\ref{finite} we obtaion the relation between dynamic $\alpha_{inc}$ and
thermodynamic $\alpha_c$ critical couplings
\begin{equation}
\frac{\alpha_{inc}}{\alpha_c} e^{s\alpha_{c}}=\frac{2es(1-s)}{\sin(\pi s)} , \hspace{0.5cm}T=0.
\end{equation}

This result shows that $\alpha_{inc}/\alpha_{c}\sim O(1)$ for $0<s \leq 1$ but different in general.

However, this result only applies if the use of the Born-Markov approximation is valid and, except
possibly at extremely weak coupling,\cite{shnirmannoise} this is not a good approximation for sub-Ohmic baths.
The Born-Markov approximation fails for sub-Ohmic baths due to the presence of low frequency modes which cause
large correlation times and strong memory effects. It is generally recognised that non-Markovian effects lead to
stronger decoherence than that described by the simple Markov rate, and at present there is much
discussion of non-Markovian dynamics in the context of qubit decoherence rates\cite{shnirman2,divencenzo,shiokawa,falci}.

As we mentioned in section \ref{var}, our method is based on a simple and intuitive variational ground-state, and we have ignored the dynamical effects of the perturbations in our discussion of the transition between coherent and incoherent phases of this ground state. The existence of a finite $\k$ as the signature for the coherent phase can be thought of as a thermodynamic criterion for the coherent phase, and in light of the discussion above, the critical couplings we have deduced from thermodynamical considerations can be different from those deduced from dynamics.

\section{Corrections and failures of the variational ground-state}
\label{correction}

The determination of $\k$ relies on the minimisation of the free energy bound,
$A_{B}$, given in Eq.(\ref{ab}).   In this section we estimate the
higher-order corrections to this free energy. We have already shown that the first
order term in powers of the perturbation vanishes and so the first corrections
are given by second-order terms. The second order term in the
Bogoliubov-Feynman bound on the free energy\cite{feynmanstat} is given by,

\begin{equation}
A_{B}^{(2)}=-\frac{1}{2}\left\langle\int^{\beta}_{0}e^{W\hat{H}_{0}}\hat{V}e^{-W\hat{H}_{0}}\hat{V}d\,W\right\rangle_{0}
\label{A2}
\end{equation}

The calculation of the second order contribution to the free energy is outlined
in appendix \ref{app1}. For weak coupling at $T=0$,
%expand all the perturbation terms to first order in $\alpha$ and
we find that the contribution to
the free energy is small,
$\frac{A_{B}^{(2)}}{A_{B}}\sim O[(\frac{\k}{\omega_{c}})^{s}]$ for $\omega_c \rightarrow \infty$.
Therefore we expect that our calculations
based on the minimisation of $A_{B}$ to be accurate in the weak coupling
regime.
For stronger coupling the corrections have to be calculated numerically, and again we find that corrections to $A_{B}$ are small when $\omega_{c}$ is much larger than all other energy scales.

Another potential weakness of the method is that the variational ansatz may not
be a particularly good guess
at the true ground state in the first place. We can in fact demonstrate some
cases where the variational solution is
sub-optimal. For simplicity, we shall show this by considering a spin-boson
Hamiltonian with only one bath mode.

We call bosonic modes adiabatic if the frequency of such modes is much larger than TLS frequency
 $\omega_b \gg K$, because these
modes can follow the TLS adiabatically.
The Silbey-Harris approach is accurate in treating these adiabatic modes as well as
being exact in the $K=0$ localized state.
Now we turn to the opposite situation, the anti-adiabatic case, where $K \gg
\omega_b$.

We introduce a different variational wavefunction for the TLS in the basis
of $\z$. It is given by,

\begin{equation}
\ket{\Psi}=\frac{1}{\sqrt{1+|\phi|^{2}}}\left(\,
\begin{array}{ll}
1&\\
\phi&\end{array}\right)
\label{phi}
\end{equation}

where the number $\phi$ is a real variational parameter to be determined and
$\phi=0$ corresponds to $\ket{\uparrow}$.
Notice that in this ansatz we allow parity symmetry (up and down direction for
the spin) to be broken unlike in the Silbey-Harris approach.
We fix the TLS in the variational state, and this gives us an effective
Hamiltonian for the bath mode\footnote{This is obtained by calculating
$\bra{\Psi}H_{sb}\ket{\Psi}$.} given by,
\begin{equation}
H_{eff}=-\frac{2K\phi}{1+\phi^{2}}+g\left(\frac{1-\phi^{2}}{1+\phi^{2}}\right)(a+a^{\dagger})+\omega\,a^{\dagger}\,a
\end{equation}

This is an example of an independent boson Hamiltonian and can be diagonalised
exactly\cite{mahan}. The resultant ground state energy is given by,

\begin{equation}
E_{g.s.}=-\frac{2K\phi}{1+\phi^{2}}-\frac{g^{2}}{\omega}\left(\frac{1-\phi^{2}}{1+\phi^{2}}\right)^{2}
\label{egs0}
\end{equation}
and we minimise this energy with respect to $\phi$ to find the optimal ground
state wavefunction. The result is that the optimal value of $\phi$ is given by,
\begin{equation}
\phi_{\pm}=\frac{2g^{2}}{\omega K}\pm\sqrt{\left(\frac{2g^{2}}{\omega
K}\right)-1}
\end{equation}
This result shows that the ground state spin is a linear combination of the
localised and delocalised states c.f. the variational method where the spin
ground state is a purely delocalised or localised state. The part of the
wavefunction corresponding to the localised state gains a displacement energy,
whilst the tunneling energy is reduced as the tunneling part of the
wavefunction has a reduced weight due to the normalisation of the wavefunction.

We notice that the important parameter here is $g^2/(\omega K)$. When
$g^2/(\omega K) \gg 1$, what we can call the strong coupling case,
the parity breaking (i.e. $\phi_{+} \gg 1$) is large. Since we assumed $K \gg
\omega$, the strong coupling case implies
that $g \gg \omega$. We will now show that when this strong coupling condition is met, the Silbey-Harris method gives a sub-optimal groundstate.

For $2g^{2}/\omega\,K\gg1$, which corresponds to either
strong coupling or a very low frequency bath mode, the ground state energy of Eq.(\ref{egs0}) is,

\begin{equation}
E_{g.s}\approx-\frac{g^{2}}{\omega}\left[1+\left(\frac{\omega\,K}{2g^{2}}\right)^{2}\right].
\end{equation}
The corresponding bound found using the Silbey-Harris method at $T=0$ is,
\begin{equation}
A_{B}\approx-K\exp\left[-\frac{g^2}{2\k^{2}}\right]-\frac{g^{2}}{\k}
\end{equation}
There is a large region of parameters that the ansatz of Eq.\ref{phi} has lower
ground state energy
than the Silbey-Harris ansatz. In particular,
if we set $g^{2}/K^{2}\ll1$, then if $\omega$ is sufficiently small so that
$\omega \ll g$, we find that,
\begin{equation}
A_{B}\approx-K-\frac{g^{2}}{K}\gg E_{g.s}.
\end{equation}
Therefore, when these conditions are satisfied, the Silbey-Harris variational
method is sub-optimal. For constant coupling, $g$, we always enter this breakdown regime as $\omega\rightarrow 0$.
However, since the coupling constant $g(\omega)$ can be frequency dependent, there
can be (and are) many situations when weak coupling
is valid as $\omega \rightarrow 0$, provided $g(\omega)$ vanishes quickly enough to maintain $g(\omega) \ll \omega$.

These results show that the coherent state found by the Silbey-Harris method can be sub-optimal for baths with finite couplings between the TLS and low frequency modes. In appendix \ref{app2} we highlight this by comparing the variational ground state given by (\ref{phi}) and the Silbey-Harris state in the limit $s\rightarrow0$.

\section{Discussion}
\label{disc}
In section \ref{var} we stated that the physical picture behind Silbey-Harris
approach is that the tunneling particle
drags along a cloud of displaced oscillators as it tunnels between the wells.
For modes with frequencies much larger than the tunneling frequency we expect
this adiabatic approximation to work well. The complications arise in this
problem due to the presence of low frequency modes in the bath, especially in
the sub-ohmic problem. These non-adiabatic modes cannot follow the tunneling
motion and need to be treated separately from the adiabatic modes.

 If we try and treat all modes with the same adiabatic approximation and set
$\f=\g$, then it can be seen that the integral in Eq.(\ref{F})
diverges in the infra-red and always leads to $\k=0$ i.e. no coherent
oscillations. This complete suppression of tunneling for the sub-ohmic bath
was also obtained by Leggett et al\cite{leggett1} using the technique of
adiabatic renormalisation.

However, the variational method goes beyond the adiabatic approximation and
finds solutions with finite $\k$.
The appearance of a finite $\k$ can be traced back to the free energy bound we
calculated in Eq.(\ref{ab}) and it is shown
explicitly in Eq.(\ref{f1}). There are two competing processes, the
choice $\f=\g$ maximises the second term,
the dressing/displacement energy. However, for sub-ohmic baths this always
renormalises $\k$ to zero and thus incurs an energy penalty. Eq.(\ref{f1}) is a non-linear
function of $\alpha,K,T$ and which process dominates depends sensitively on
these parameters. When $\alpha<\alpha_{c}(T)$
it is energetically favourable to have a finite $\k$.

For $\alpha<\alpha_{c}$ and $T=0$, we see from Eq.(\ref{f}) that the variational method has
loosely separated the bath modes into two distinct sets. Modes with
$\omega>2\k$ respond adiabatically to the tunneling motion i.e. have
$\f\approx\g$. Non-adiabatic modes with $\omega<2\k$ couple more weakly to the TLS, with coupling strength
$\f\approx \g \frac{\omega_l}{2\k}$ as $\omega_l \rightarrow 0$.

 This vanishing of the coupling at low frequencies prevents the infra-red
divergence in Eq.(\ref{F}) by fixing an effective cut-off at $2\k\tanh(\k\beta)$.
In this method, the free energy minimisation naturally determines the cut-off
for the mode elimination, unlike in the adiabatic renormalisation
scheme\cite{leggett1,kehrein}. We also note that while the non-adiabatic modes
decouple from dressing the particle, they have not disappeared; they give the
dominant contribution to the perturbation term $\hat{V}_{0}$, Eq.(\ref{dep}), and can cause
significant dynamical effects.

The variational method also predicts interesting behaviour for $\k (T)$ at low
temperatures. As we demonstrated in section \ref{T}, $\k(T)$ initially
increases with temperature and this behaviour can be seen to arise from the
non-adiabatic modes. Interestingly, we find that the renormalisation of $\k(T)$
due to the non-adiabatic modes actually decreases at finite temperatures. We
would normally expect that as the temperature is increased, the occupation of
low frequency oscillators would increase, and this  should lead to increased
renormalisation through the hyperbolic cotangent factor in Eq.(\ref{F}).
However, from Eq.(\ref{f}) we see that the dressing due to modes with $k_{B}T<\omega<2\k$ decreases with temperature, and this decoupling leads to an overall reduction in the renormalisation of $\k$ due to these non-adiabatic modes.

The dressing parameters for the adiabatic modes are effectively independent of
temperature, and so when they are thermally excited they always renormalise
$\k$ towards zero. At low temperatures when there are almost no adiabatic modes
excited, the reduction in the renormalisation due to non-adiabatic modes leads
to the increase of $\k$ with temperature. This low temperature reduction in the renormalisation of $\k$ also gives a natural explanation for the re-entrance of finite $\k$ at finite temperatures for systems with $\alpha>\alpha_{c}$. At high enough temperatures, the
renormalisation due to excited adiabatic modes always dominates and $\k$ then decreases
until it goes discontinuously to zero at $T^{*}$.

  The exact point at which the adiabatic modes halt the increase in $\k$
depends sensitively on the relative weight of adiabatic and non-adiabatic modes
and thus depends on the spectrum of the bath. However, we can still estimate
where the maximum occurs. From the discussion above, the turning point occurs
at the temperature at which the adiabatic modes begin to be excited. This
occurs approximately at a temperature $K_{B}T_{max}\sim\k(T_{max})$.

There have been several other recent treatments of the sub-ohmic problem\cite{kehrein,shnirmannoise,bulla} and we find that this simple variational method is consistent with several of the main results.
The flow equation analysis of Kehrein and Mielke\cite{kehrein} also
showed that a coherent phase
 exists for the sub-ohmic model.
On the basis of the well-known connection between spin-boson model and
Ising model in statistical mechanics\cite{spohn}, the coherent phase, corresponding to the high-temperature
disordered phase of the Ising model, is expected to exist. Many results of Ref.\cite{kehrein} are in fact
consistent with ours, including a qualitative
prediction of the rise in $\k(T)$ at low temperatures and the discontinuous
transition at zero temperature.

It is important to remember that the transition in the spin-boson
model as a function of coupling constant
$\alpha$ can be related
to the transition in an infinite one-dimensional Ising model with long-range
interactions as a function of temperature, only when $T=0$ in the spin-boson model.
Therefore
comparison of the nature of the transition (1st-order or 2nd-order type) is
limited to $T=0K$.
In this paper we are mostly concerned with the transition of spin-boson model at
finite temperature, with
several parameters describing the bath ($\alpha, \omega_s,\omega_c$). Yet the
comparison with the results known
for the Ising model with $1/r^{1+s}$ interactions indicates that higher-order corrections to Silbey-Harris
ansatz should be necessary
to describe the close proximity of the transition, since for $s>0$ the
transition in the Ising model
is of 2nd-order.\cite{fisher}

The numerical renormalisation group analysis by Bulla, Tong and
Vojta\cite{bulla}  found that the system is localized at $s=0$,
and their perturbative RG results suggest that for $s>0$, the transition is
continuous as a function of $\alpha$. As our method
is based on a variational ansatz, we
cannot make any strong statement about
the exact nature of the transition. As we noticed in section\ref{o} there are
several parameters
which describe the bath, and the transition may depend on the constraints
between parameters imposed and assumptions used
in the mappings to other models.

Shnirman, Makhlin and Schon\cite{shnirmannoise} have also demonstrated that
coherent oscillations are possible in the sub-ohmic model,
but their work focusses on calculating the dephasing and relaxation times of
the dynamics rather than renormalisation effects.
In contrast to Bulla, Tong and Vojta, their diagrammatic approach predicts that the TLS can be coherent at $T=0$ and
$s=0$.
As we discussed, differences between thermodynamic and dynamic properties are
expected for the spin-boson model
with a sub-Ohmic bath, and further understanding of these questions is desirable.

\section{Conclusions}

We have studied the sub-ohmic spin boson model using the intuitive
variational method of Silbey and Harris.\cite{silbey} This method
has allowed us to reproduce a number of previously known results about the
coherent sub-Ohmic model, but without having to make lengthy or unduly
complicated calculations. With this in mind, we note that this method may be
useful for a first look at different types of environment for which there is
some question about the existence of a coherent phase.

For the $T=0$ Sub-ohmic spin boson model, we have shown that coherent
oscillations exist if $\alpha$ is below a critical coupling, $\alpha_{c}$ which
we have explicitly calculated in Eq.(\ref{zero}). When this condition is met, the
renormalised tunneling matrix element $\k$ satisfies, $Ke^{\frac{1}{s-1}}\leq
\k\leq K$ and undergoes a discontinuous transition to $\k=0$ as
$\alpha\rightarrow\alpha_{c}$.

We have also presented new numerical results which show the dependence of
$\k(T,\alpha)$ on temperature and coupling strength. We have shown that $\k(T)$
has a non-trivial dependence on temperature, initially rising to a maximum
value and then decreasing
to a discontinuous transition at a critical temperature $T^{*}$.
We were able to show that this behaviour arises from the temperature dependence
of the effective dressing parameters $\{\f\}$(\ref{f}), and we have highlighted
the natural separation in this method of adiabatic modes ($\omega>2\k$) and
non-adiabatic modes $(\omega<2\k)$. Our numerical study of this theory also found a new phenomena, a re-entrant coherent phase that exists at finite temperatures for systems with $\alpha>\alpha_{c}$ if $\alpha$ is sufficiently close to the critical coupling.

Importantly, we showed that dynamical and thermodynamic criteria for the
transition
are different and sensitive to non-adiabatic modes. We also discussed several
limitations
of the description of the spin-boson model by an equilibrium bath characterized by
the spectral function $J(\omega)$.

We would like to thank P. B. Littlewood for useful discussions.

%------------------------------------------------------------------------------------------

\appendix
\section{Calculation of second-order terms for the Free-energy bound}
\label{app1}
In section \ref{correction} we discussed the size of contributions to the free energy from higher order terms in equation (\ref{ab}). In this Appendix we outline the calculation of the lowest order correction terms to the free energy bound. The fist corrections are second-order in the perturbations (\ref{V}) and given are by,
\begin{equation}
A_{B}^{(2)}=-\frac{1}{2}\left\langle\int^{\beta}_{0}e^{W\hat{H}_{0}}\hat{V}e^{-W\hat{H}_{0}}\hat{V}d\,W\right\rangle_{0}.
\label{f2}
\end{equation}
The perturbation terms are shown in equation (\ref{V}) and the Hamiltonian
$\hat{H}_{0}$ is defined in equation (\ref{ho}).
The average is explicitly given by,
\begin{equation}
\langle
A\rangle_{0}=\frac{\textrm{Tr}\exp(-\beta\,H_{0})A}{\textrm{Tr}\exp(-\beta\,H_{0})}.
\end{equation}

Each perturbation term is a product of a spin operator and a bath operator. As
the thermal density matrix corresponding to $H_{0}$ is also separable into spin
and bath parts, we can  calculate each term in equation (\ref{f2}) as the
product,

\begin{equation}
\int^{\beta}_{0}d\,W\left\langle\,e^{W\hat{H}^{s}_{0}}\,\hat{V_{s}}e^{-W\hat{H}^{s}_{0}}\,\hat{V_{s}}\right\rangle_{s}\left\langle\,e^{W\hat{H}^{b}_{0}}\,\hat{V_{b}}e^{-W\hat{H}^{b}_{0}}\,\hat{V_{b}}\right\rangle_{b}
\label{prod}
\end{equation}

here $s$ refers to the spin part of $H_{0}$ and $b$ is the bath part. Before
discussing these factors,
it is useful to re-write the perturbations in terms of the spin components $x,y,z$ instead of the raising and lower operators. This gives the perturbations as,
\begin{eqnarray}
\hat{V}_{0}&=&\s\,(\g-\f)\,(\a +\adag)\z\\
\hat{V}_{1}&=&\k\left[\cosh\{2\s\,\f\,\freq^{-1}(\a -\adag)\}-1\right]\x\\
\hat{V}_{2}&=&-i\k\sinh\left[2\s\,\f\,\freq^{-1}(\a-\adag)\right]\hat{\sigma}_{y}\label{VV}
\end{eqnarray}

\subsection{Spin part}
The spin factor is of the general form,
\begin{equation}
I^{ijk}_{s}=\left\langle
e^{W\k\sigma_{i}}\,\sigma_{j}e^{-W\k\sigma_{i}}\,\sigma_{k}\right\rangle_{s}.
\label{spinfactor}
\end{equation}
where $i=x,y,z$. The exponentiated spins can be written\cite{landauQM},
\begin{equation}
\exp(\theta\sigma_{i})=\cosh(\theta)+\sinh(\theta)\sigma_{i}
\end{equation}
 and using this and the pauli spin algebra, one can derive the general
relationship,
\begin{eqnarray}
e^{\theta\sigma_{i}}\,\sigma_{j}\,e^{-\theta\sigma_{i}}&=&\cosh(2\theta)\sigma_{j}+i\epsilon_{ijk}\sinh(2\theta)\sigma_{k}\nonumber\\
&-&2\sinh^{2}(\theta)\delta_{ij}\sigma_{j}.
\end{eqnarray}

Substituting this into (\ref{spinfactor}) we get,
\begin{eqnarray}
I^{ijk}_{s}&=&\cosh(2\k\,W)\langle\sigma_{j}\sigma_{k}\rangle_{s}\nonumber\\
&+&i\,\epsilon_{ijl}\sinh(2\k\,W)\langle\sigma_{l}\sigma_{k}\rangle_{s}\nonumber\\
&-&2\sinh^{2}(\k\,W)\delta_{ij}\langle\sigma_{j}\sigma_{k}\rangle_{s}
\end{eqnarray}

and finally, all the spin factors can be calculated using,
\begin{eqnarray}
\langle\sigma_{x}\rangle_{s}&=&-\tanh(\k\beta)\\
\langle\sigma_{y}\rangle_{s}&=&\langle\sigma_{z}\rangle_{s}=0.
\end{eqnarray}

\subsection{Bath factors}
If we define the operator $a_{l}(W)=\exp(W\,H_{b})a_{l}\exp(-W\,H_{b})$, then
the bath terms contain only averages of the form,
\begin{equation}
I_{b}=\big\langle V_{i}(W)V_{j}\big\rangle_{b}
\end{equation}
where the $V_{i}$ are the bath parts of the perturbation terms defined in
equation (\ref{VV}). To continue we need to calculate these expectation values.
For the terms involving products of $V_{1,2}$ the following theorem is very
useful. If the operators $A$ and $B$ are linear in the co-ordinates or momenta
of an oscillator, then it can be shown\cite{ashcroft}

\begin{equation}
\langle\,e^{A}e^{B}\rangle_{b}=e^{\frac{1}{2}\left[\langle\,A^{2}\rangle_{b}+\langle\,B^{2}\rangle_{b}+2\langle\,AB\rangle_{b}\right]}
\end{equation}
for example, if we define $\Delta(W)=2\sum_{l}\f\omega_{l}^{-1}(\a(W)-\adag(w))$, then

\begin{eqnarray}
\langle
V_{2}V_{2}\rangle_{b}&=&-\frac{K^{2}}{4}\left\langle(e^{\Delta(W)}-e^{-\Delta(W)})(e^{\Delta(0)}-e^{-\Delta(0)})\right\rangle\nonumber\\
&=&-\frac{K^{2}}{2}e^{\langle\Delta^{2}\rangle}\left[e^{\langle\Delta(W)\Delta(0)\rangle}-e^{-\langle\Delta(W)\Delta(0)\rangle}\right]\nonumber\\
&=&-\k^{2}\sinh(\gamma(W))
\end{eqnarray}

where $\gamma(W)$ is given by,

\begin{eqnarray}
\gamma(W)&=&\langle\Delta(W)\Delta(0)\rangle\\
&=&-4\sum_{l}\f^{2}\omega^{2}_{l}\left[e^{\,\omega_{l}W}n_{l}+e^{-\omega_{l}W}(n_{l}+1)\right]\nonumber
\end{eqnarray}
and we have used $\k=K\exp(\frac{1}{2}\langle\Delta(0)^2 \rangle)$.

Combining these results with the spin factors, we calculate that the second-order
contribution to the free energy is,
\begin{widetext}
\begin{eqnarray}
&2A_{B}^{(2)}&=-\k^{2}\int_{0}^{\beta}d\,W\,[\cosh(\gamma(W))-1]\label{ch}\\
&+&\k^{2}\int_{0}^{\beta}d\,W\,\sinh(\gamma(W))\left[\,\cosh(2\k\,W)-\sinh(2\k\,W)\tanh(\k\beta)\right]\\\label{sh}
&-&\int_{0}^{\beta}d\,W\left\langle\,V_{2}(W)V_{0}\right\rangle\left[\,\sinh(2\k\,W)-\cosh(2\k\,W)\tanh(\k\beta)\right]\\
&+&\int_{0}^{\beta}d\,W\left\langle\,V_{0}(W)V_{2}\right\rangle\left[\,\sinh(2\k\,W)-\cosh(2\k\,W)\tanh(\k\beta)\right]\\
&-&\s(\g-\f)^{2}\int^{\beta}_{0}dW\left(e^{W\omega_{l}}\tilde{n}_{l}+(\tilde{n}_{l}+1)e^{-W\omega_{l}}\right)\left(\cosh(2\k\,W)-\tanh{\k\beta}\sinh(2\k\,W)\right).\label{zz}
\label{f2t}
\end{eqnarray}
\end{widetext}

Note that the free energy correction, $A_{B}^{(2)}$, is stated for the case of finite $\k$. For $\k=0$, $A_{B}^{(2)}$ is identically zero.

\subsection{Second-order terms at $T=0$}
At $T=0$ we can calculate all expectation values in (\ref{f2t}) explicitly and the free energy correction takes the form,
\begin{eqnarray}
A_{B}^{(2)}&=&-\frac{\k^{2}}{2}\int_{0}^{\infty}d\,W\,[\cosh(\gamma(W))-1]\label{ch0}\\
&+&\frac{\k^{2}}{2}\int_{0}^{\infty}d\,W\,\sinh(\gamma(W))e^{-2\k\,W}\label{sh0}\\
&-&6\k^{2}\sum_{l}\frac{\g^{2}}{(\omega_{l}+2\k)^{3}}\label{k0}.
\end{eqnarray}

We will show that the typical size of the correction term is small compared to the main free energy in the limit of large $\omega_{c}$, which is the normal situation in this model. For the term (\ref{k0}) we get a contribution of,

\begin{eqnarray}
-6\k^{2}\sum_{l}\frac{\g^{2}}{(\omega_{l}+2\k)^{3}}&\approx&-\frac{\alpha\omega_{s}}{2}\left(\frac{2\k}{\omega_{s}}\right)^{s}
\left( \frac{1}{1+s} +\frac{1}{s-2} \right),\nonumber\\
\label{secondorderT=0}
\end{eqnarray}
where we have introduced the spectral function and approximately calculated the
integral. The other two terms,(\ref{ch0}) and(\ref{sh0}), cannot be evaluated in a simple analytical form, but we note that as $\omega_{c}\rightarrow\infty$ these terms give finite contributions if $s<1$.

The main part of the free energy $A_{B}$ is given by,
\begin{eqnarray}
A_{B}&=&-\k+\s(\f^{2} -2\,\f\,\g)\label{f1}\\
&=&-\k
-\frac{\alpha\omega_{s}^{1-s}}{2}\int^{\omega_{c}}_{0}\frac{(\omega+4\k)\omega^{s}d\omega}{(\omega+2\k)^{2}}
\end{eqnarray}
where again, we have used the spectral function to convert the sum into an
integral. Under assumption  $\k \ll \omega_c$, the leading
term in $\omega_c$ of  $A_{B}$ is
\begin{equation}
A^{B}\approx-\frac{\alpha\omega_{s}}{2s}\left(\frac{\omega_{c}}{\omega_{s}}\right)^{s}.
\label{ABestimate}
\end{equation}
Comparing this to the second order correction $A_{B}^{(2)}$, we see that
corrections due to the term given by (\ref{k0}) are small, and are controlled by the small parameter
$(\frac{\k}{\omega_c})^s$ for $s>0$. As we let $\omega_{c}$ become large, the corrections from terms (\ref{sh0}) and (\ref{ch0}) tend to a finite value, whilst $A_{B}$ grows as $\omega_{c}^{s}$. Therefore, the relative correction from (\ref{sh0}) and (\ref{ch0}) becomes small in this limit.
However, these and higher order perturbations
may still be relevant in the proximity of the coherent-incoherent transition as $\k\rightarrow 0$.

\section{Alternative variational treatment of spin-boson problem}
\label{app2}
In this section we give a treatment of the sub-Ohmic spin-boson model using the
alternative variational solution
given in section \ref{correction}. In the anti-adiabatic or non-adiabatic
situation of $K \gg \omega$, TLS
can be thought as creating effective potential for bosonic mode.
As before we write a variational state for the spin,
\begin{equation}
\ket{\Psi}=\frac{1}{\sqrt{1+|\phi|^{2}}}\left(\,
\begin{array}{ll}
1&\\
\phi&\end{array}\right)
\label{phi-again}
\end{equation}
We then calculate $\bra{\Psi}H_{sb}\ket{\Psi}$ to get the effective Hamiltonian
for the bath modes. This is given by,
\begin{equation}
H_{eff}=-\frac{2K\phi}{1+\phi^{2}}+\left(\frac{1-\phi^{2}}{1+\phi^{2}}\right)\sum_{l}g_{l}(a+a^{\dagger})+\sum_{l}\omega_{l}\,a^{\dagger}\,a
\end{equation}
The first tunneling term is minimized for real $\phi$, so that $\phi$ is chosen
to be real although in general complex.

Again, this is a set of independent boson Hamiltonians and the energy of the
variational ground state is given by,

\begin{equation}
E_{g.s.}=-\frac{2K\phi}{1+\phi^{2}}-\left(\frac{1-\phi^{2}}{1+\phi^{2}}\right)^{2}\sum_{l}\frac{g_{l}^{2}}{\omega_{l}}\label{egs}
\end{equation}
The sum over the bath couplings can be explicitly calculated by substituting
the spectral function into the sum to get,
\begin{eqnarray}
\sum_{l}\frac{g_{l}^{2}}{\omega_{l}}&=&\frac{\alpha\omega_{s}^{1-s}}{2}\int^{\omega_{c}}_{0}\omega^{s-1}\,d\omega\\
&=&\frac{\alpha\omega_{s}}{2s}\left(\frac{\omega_{c}}{\omega_{s}}\right)^{s}
\end{eqnarray}
Looking at the ground state energy (\ref{egs}) we see that as $s\rightarrow 0$,
the static displacement energy of the oscillators (given by the second term of
Eq.(\ref{egs})) diverges and becomes the dominant term for any non-zero coupling.
Minimising the free energy w.r.t $\phi$ we always find that $\phi=0$ (or
$\phi=\infty$) and therefore the particle is always localised for any non-zero
coupling at $s=0$.
This is due to the fact that these soft modes have no resistance to the
static force due to the spin in the limit $\omega_{l}\rightarrow 0$.

For $s=0$, the Silbey-Harris variational method predicts a coherent phase with finite $\k$ for sufficiently weak coupling. The free energy of this state is,
\begin{eqnarray}
A_{B}&=&-\k-\frac{\alpha\omega_{s}}{2}\int^{\omega_{c}}_{0}\frac{(\omega+4\k)d\omega}{(\omega+2\k)^{2}}\\
&\approx&-\k-\alpha\omega_{s}\ln\left[\frac{\omega_{c}}{2\k}\right].
\end{eqnarray}
Comparing the energy of this coherent ground state to the energy of the
localised ground state,
we see that for $s=0$ and $\omega_c=\infty$ the localised state is lower in energy for any non-zero
coupling between the bath and the TLS. The coherent state is therefore never favourable when $s=0$ and the finite $\k$ found by the variational method is an artefact of the method.
This artefact occurs due to the divergence of the static displacement energy of the oscillators
(singular limit for $\omega_c \rightarrow \infty$), which causes problems with the free energy minimisation we use to determine $\f$, $\k$ etc.  Notice though that $\k =0$ is also a solution of the self-consistent Eq.(\ref{F}), and so the Silbey-Harris method can correctly describe the $s=0$ state if we ignore the sub-optimal solution with $\k>0$.

To summarise, the divergence of the static displacement energy of the oscillators for $s \leq
0$  implies localization
in the ground state and dramatic differences between thermodynamic and dynamic
properties. Such differences due to
non-adiabatic modes can also be seen for $s>0$.

\bibliography{decoherence}% Produces the bibliography via BibTeX.

\end{document}